\documentclass[prl,aps,twocolumn,groupedaddress,showpacs,floatfix]{revtex4-1}
\usepackage{amsmath,amssymb,multirow,epsfig,bm,mciteplus}

\newcommand{\etal}{{\it et al.,\;}}
\newcommand{\beq}{\begin{equation}}
\newcommand{\eeq}{\end{equation}}
\newcommand{\bea}{\begin{eqnarray}}
\newcommand{\eea}{\end{eqnarray}}

\newcommand{\nn}{\nonumber}
\newcommand{\benn}{\begin{displaymath}}
\newcommand{\eenn}{\end{displaymath}}

\begin{document}

\title{Isovector Giant Dipole Resonance from the 3D Time-Dependent Density Functional Theory for Superfluid Nuclei}

\author{I. Stetcu$^1$, A. Bulgac$^1$, P. Magierski$^2$, and K.J. Roche$^{3,1}$ }

\affiliation{$^1$Department of Physics, University of Washington, Seattle, WA 98195--1560, USA}
\affiliation{$^2$Faculty of Physics, Warsaw University of Technology, ulica Koszykowa 75, 00-662 Warsaw, POLAND }
\affiliation{$^3$ Pacific Northwest National Laboratory, Richland, WA 99352, USA}

\begin{abstract}
A fully symmetry unrestricted Time-Dependent Density Functional Theory extended to include pairing correlations is used to calculate properties of the isovector giant dipole resonances of the deformed open-shell nuclei $^{172}$Yb (axially deformed), $^{188}$Os (triaxially deformed), and $^{238}$U (axially deformed), and to demonstrate good agreement with experimental data on nuclear photo-absorption cross-sections for two different Skyrme force parametrizations of the energy density functional: SkP and SLy4.
\end{abstract}

\date{\today}


\maketitle

The isovector giant dipole resonance (GDR) is perhaps the simplest example of a nuclear collective motion of all the protons against all the neutrons. Since its observation in the photo-absorption cross section \cite{bothe}, it has been intensively studied as it combines several challenging aspects of the physics of the atomic nucleus \cite{levinger,harakeh}. Even though GDR is practically harmonic in character,  it is not an adiabatic collective mode and various damping mechanisms of the collective energy are at work  \cite{bertsch}. In the models of Migdal \cite{migdal}, Goldhaber-Teller \cite{GT},and Steinwedel-Jensen \cite{SJ} GDR is described as the relative motion of two fluids, either compressible or incompressible,  with neutrons and protons vibrating around a common center of mass, and the mass dependence of the excitation energy reads $A^{-1/6}$ and $A^{-1/3}$ respectively \cite{ring}.  A good estimation of the GDR vibrational frequency is $\hbar \omega \approx 80$ MeV $A^{-1/3}$ for spherical nuclei. GDR is interpreted simply as the equivalent of the zero-sound in a nuclear system and the size of the nucleus sets a constraint on the largest wavelength. In the case of  deformed nuclei, the GDR peak is split, with various frequencies revealing different principal axes of the nuclear shape. Since the GDR state is not an eigenstate of the nuclear Hamiltonian, it is characterized by a spreading width that cannot be extracted from a simple hydrodynamical approach. The spreading width in principle can be calculated within a microscopic model of the atomic nucleus. The total width of the GDR is mainly due to a couple of mechanisms:  the coupling of the GDR to complex nuclear configurations $\Gamma^{\downarrow}$, and the coupling to the continuum, leading to the escape of neutrons and protons  $\Gamma^{\uparrow}$. These two widths contribute to the total width of the GDR, $\Gamma = \Gamma^{\downarrow} + \Gamma^{\uparrow}$, and their relative contributions vary depending on the mass number $A$ and the $N/Z$ ratio.  The escape width is typically more important for light nuclei. The physical mechanisms related to $\Gamma^{\downarrow}$ may be quite complicated  and involve coupling to low energy surface vibrations, Landau damping and collisional damping \cite{bertsch}.

The theoretical approach to GDR is typically within the linear response of the nuclear system to an external probe \cite{bb} . In contemporary approaches, the description of the atomic nucleus is provided by a Density Functional Theory (DFT) \cite{dreizler} and the GDR is described within the small amplitude limit of the time-dependent version of DFT. This approach leads to the well known (Quasiparticle) Random Phase Approximation (Q)RPA that has been used with  a variety of nuclear density functionals, though only very recently for deformed systems, see Refs.  \cite{teb,terasaki,losa,peru,jt,martini,stoitsov,avogadro,maruhn,stevenson,nakatsukasa} for several representative calculations.  When applied to open-shell and, in particular, deformed nuclei QRPA requires diagonalizations of matrices of extremely large sizes and often severe truncations of the quasiparticle basis set. Truncations lead to a number of undesired features, in particular spurious states, instead of having a zero excitation energy, get admixed to transitions of physical interest. In a deformed open-shell nucleus, states corresponding to the excitation of the center of mass motion, rotation of the system, and gauge transformations arising from the breaking of the proton and neutron number conservations, all have zero-energy excitation energy (sometimes referred to as Goldstone modes). When (Q)RPA is derived, one usually allows for small violations of the Pauli principle, which have only a negligible effect when one is considering the most collective transitions. (Q)RPA formally is an approach in which the fermionic nuclear Hamiltonian is bosonized and the number of bosonic excitations allowed to exist is significantly much larger than the number of true physical excitations.  Significant difficulties arise when one tries to go beyond the (Q)RPA and consider coupling to more complex configurations, and describing  the spreading width $\Gamma^\downarrow$. In spite of the efforts of many generations of theorists and many approaches suggested so far (nuclear field theory \cite{mottelson}, boson expansion methods \cite{ring,klein}, and diagrammatic methods \cite{bertsch}), a truly satisfactory  microscopic treatment is still lacking and theorists resort to either phenomenological models, such as optical potentials (which can easily lead to overestimates of $\Gamma^\downarrow$ unless vertex corrections are accounted for as well \cite{bertsch}),  or simple prescriptions such as adding a popular, but arbitrary, energy smoothing of the (Q)RPA strength.   

Here we present an approach to GDR in open-shell nuclei based on an extension of the DFT to superfluid systems, superfluid local density approximation (SLDA)  and its further extension to time-dependent phenomena (TDSLDA), which was developed over the years and applied to a number physical systems and phenomena, see Refs. \cite{by1,yb1,slda1,yb2,by2,slda2,bf,yoon,bfm,bulgac,blr,kenny}. The time-dependent Hartree-Fock-Bogoliubov (TDHFB)  approximation has been used previously as an alternative to QRPA in spherical nuclei \cite{avez} and with a number of further approximations in deformed nuclei \cite{hashimoto,takahashi}. TDSLDA appears formally as a time-dependent self-consistent local mean-field approximation. No spurious modes are admixed to physical transitions, since the energy density functional respects all the required symmetries (translational and rotational symmetry, Galilean invariance \cite{galilean}, parity, isospin symmetry, gauge symmetry)  and no further approximations are introduced, apart from discretization errors which are under control. Isospin symmetry is broken only by the proton/neutron mass difference and by the Coulomb energy, which we treat in the Hartree approximation. We use an approximation to the normal part of the nuclear energy density functional provided by various Skyrme force parametrizations \cite{skp,sly4,bhr} and all the terms are taken into account numerically exactly. The pairing part of the energy density functional is treated as described in Ref. \cite{by1} and \cite{yb1}, when a single coupling constant is used for both protons and neutrons, and even and odd particle numbers as well,  unlike most phenomenological approaches to nuclear pairing which break isospin invariance in the pairing channel. In such approaches, see Refs. \cite{terasaki} and \cite{jt} for examples of QRPA calculations, the proton pairing coupling constant  is larger in magnitude than the neutron pairing coupling constant, $|V_p|>|V_n|$. This leads to a flagrant violation of isospin invariance in the energy density functional, the magnitude of which is clearly not due to  charge-symmetry-breaking forces. Such effective pairing couplings would be consistent with a new kind of pairing energy, that does not violate charge symmetry breaking \cite{csb}. The Coulomb interaction will lead to just an opposite relation \cite{duguet}, namely to $|V_p|<|V_n|$.  Galilean invariance \cite{galilean} requires the presence of currents in the energy density functional, and although their contribution is vanishing in the ground states of even-even nuclei, it is important for the excited states \cite{t-odd,mizuyama}.

Within TDSLDA one describes accurately the interaction in both particle-hole and particle-particle channels, the treatment is fully self-consistent, and all symmetries of the Hamiltonian are properly accounted for. We place the nuclear system on a 3D spatial lattice, needed spatial derivatives are determined using Fast Fourier Transform, and we at first determine the ground state properties within SLDA and subsequently subject the system to an external time-dependent one-body potential \cite{kenny,calvayrac,bulgac}. The Coulomb interaction is treated with particular care, so as not to introduce the potential due to image charges, which appear naturally in a spatial lattice formulation with periodic boundary conditions. Although we study the small amplitude limit, the equations we solve here are the same as for a motion of arbitrary amplitude.  The emerging equations are formally equivalent to the TDHFB approximation with local potentials,  or to the time dependent Bogoliubov-de Gennes (TDBdG) equations:
\bea
i\hbar 
\left  ( \begin{array} {c}
  \dot{u}_{k\uparrow}\\ \dot{u}_{k\downarrow}\\ 
  \dot{v}_{k\uparrow} \\ \dot{v}_{k\downarrow}\\
\end{array} \right ) =
\left ( \begin{array}{cccc}
h_{\uparrow\uparrow}  & h_{\uparrow\downarrow} &0&\Delta \\
h_{\downarrow\uparrow}  & h_{\downarrow\downarrow} &-\Delta &0\\
0&-\Delta^* &-h^*_{\uparrow\uparrow} & -h^*_{\uparrow\downarrow}\\
\Delta^*&0&-h^*_{\uparrow\downarrow} & -h^*_{\downarrow\downarrow}
\end{array} \right )  
\left  ( \begin{array} {c}
  u_{k\uparrow}\\ u_{k\downarrow}\\ 
  v_{k\uparrow}\\ v_{k\downarrow}\\
\end{array} \right ). \nn
\eea
where we have suppressed the spatial ${\bf r}$ and time coordinate and $k$ is the label of each quasiparticle wave function $[u_{k\sigma} ({\bf r},t), v_{k\sigma} ({\bf r},t)]$. where $\sigma=\uparrow,\downarrow$.  The single-particle Hamiltonian $h_{\sigma\sigma'}({\bf r},t)$ is a partial differential operator (thus local)  and $\Delta({\bf r},t)$ is a pairing field, all defined through the normal, anomalous, spin, and isospin densities and currents. The interaction with various applied external fields (spin, position and/or time-dependent) is described by including the corresponding potentials in the single-particle Hamiltonian $h_{\sigma\sigma'}({\bf r},t)$.  This approach represents a flexible tool to describe in general large amplitude nuclear motion as it contains the coupling to the continuum and between single-particle and collective degrees of freedom, since the meanfield is explicitly time-dependent. However, the later type of coupling between single-particle and collective degrees of freedom will account only for a part of the diagrams discussed in Ref. \cite{bertsch}. If the external field will act along one of the symmetry axes for example, only collective oscillations along that axis will be excited, but not in the perpendicular direction. Although the TDSLDA is designed to provide the values of one-body densities, a simple modification of the variational principle extends the approach to two body observables, for example, particle number fluctuation \cite{balian1,balian2,balian3,broomfield1,broomfield2}.  Another straightforward extension of the formalism leads to the stochastic TDSLDA, with hopping between various time-dependent meanfields, and can account for dissipation \cite{stoch}.

The most convenient quantity for studying the GDR is the strength distribution: $ S(E)=\sum_{\nu}|\langle \nu|\hat{F}|0\rangle |^{2} \delta( E - E_{\nu})$, where $| \nu \rangle $ are nuclear eigenstates corresponding to energies $E_{\nu}$. In our case the operator $\hat{F} $ is a sum of two operators depending on neutron and proton coordinates, respectively: $F_\tau({\bf r}) =  N_\tau\sin( {\bf k}\cdot{\bf r} _\tau )/|{\bf k}| $, where $\tau=n,\; p$, $|{\bf k}|=2\pi /L$, $L$ is a lattice size of the box, $N_{n}=-Z, N_{p}=N$ are neutron and proton numbers, respectively and $A=N+Z$. The operator $F_\tau({\bf r})$  generates all the odd multipoles, but the predominant contribution comes from the dipole mode. In practice, depending on the symmetry of the ground state, we calculate the response of the nucleus to different external fields. Thus, for a spherical nucleus, we can chose ${\bf r}_p$ and ${\bf r}_n$ along any direction, for an axially deformed nucleus we compute  two different responses with ${\bf r}_p$ and ${\bf  r}_n$ along two symmetry axes, while for a triaxial nucleus we compute the responses along three principal axes. Within our formalism, the external perturbation is added to the Hamiltonian $h_{\tau,\sigma\sigma}({\bf r},t)\Rightarrow h_{\tau,\sigma\sigma}({\bf r},t) + F_\tau({\bf r}) f(t)$ and switched on adiabatically, where $f(t)=C \exp[-(t-10)^2/2]$ (time in units of fm/c)  and $C$ defines the intensity of the perturbation and which has to be kept sufficiently small to stay in the linear response regime, but large enough to excite the modes of interest.  The amount of energy deposited into a nucleus was in the range 45-50 MeV. For such perturbations, which imply harmonicity of the excitation, the response function to the external perturbation $\hat{F} f(t)$ is given by \cite{calvayrac}: $S(\omega) = {\mathrm{Im}} \{\delta F(\omega)/[\pi f(\omega)]\}$,  where $\delta F(\omega)$ is the Fourier transform of the fluctuation of the expectation value: $ \delta F(t) = \langle \hat{F}\rangle_{t} - \langle \hat{F}\rangle_{0} = \int d^{3}r\delta \rho({\bf r},t) F({\bf r})$. The extraction procedure of the strength function is simple to state: obtain the self-consistent stationary solution using the stationary SLDA solver to high precision so the spurious contributions due to global excitations (center-of-mass motion, rotations for deformed nuclei, proton and neutron pairing Goldstone modes) are decoupled, and subsequently use this solution as input for the TDSLDA code where it is perturbed by the external field and evolved for a period of time, $T$. The energy resolution due to taking the Fourier transform in the finite window $T$ is $ \delta E=2\pi\hbar /T$, and so it is important to perform propagation to sufficiently large $T$ (here $T\approx 1600$ fm/c and $\delta E\approx 0.8$ MeV).  We have used a cubic lattice box with $L=32.5$ fm on a side and a lattice constant 1.25 fm, which allows us to obtain the ground-state energy with an accuracy of a few tens of kilo-electron-volts for a number of spherical light open-shell nuclei.  The TDSLDA equations were integrated in time using a fifth order multistep method with a  time step $\Delta t\approx 0.12$ fm/c chosen to maintain a $10^{-7}$ relative accuracy, see Refs. \cite{kenny,bulgac}. The changes in proton and neutron numbers during the time evolution change are $|\Delta Z|<10^{-4}$ and $|\Delta N| < 10^{-4}$ respectively. TDSLDA, when the external perturbation does not change the particle number, as in the GDR case, the particle number in the time-dependent solution is exactly conserved, and thus all the excited modes have exactly the same average proton and neutron particle number as the ground state. The method recently suggested in Ref. \cite{avogadro}, with the technical improvements described in Ref. \cite{stoitsov},  has the promise to become the leading approach for QRPA calculations in open-shell nuclei, because of the simplicity of its numerical implementation and its accuracy. The solutions of the QRPA however are harder to interpret \cite{teb}, since for a given nucleus these equations will provide solutions for $\Delta N=0, \; \Delta Z=0$,  as well as pairing vibration type transitions with $\Delta N=\pm 2$ and/or $\Delta Z=\pm2$ respectively.  TDSLDA has been applied so far to describe excitations with no change in particle number $\Delta N =0$ due to time-dependent external fields which do not change the particle number \cite{bulgac,blr}, as well as excitations with $\Delta N \neq 0$ \cite{yoon} in the case of more complicated time-dependent external probes.
 
\begin{figure}
\includegraphics*[width=0.45\textwidth]{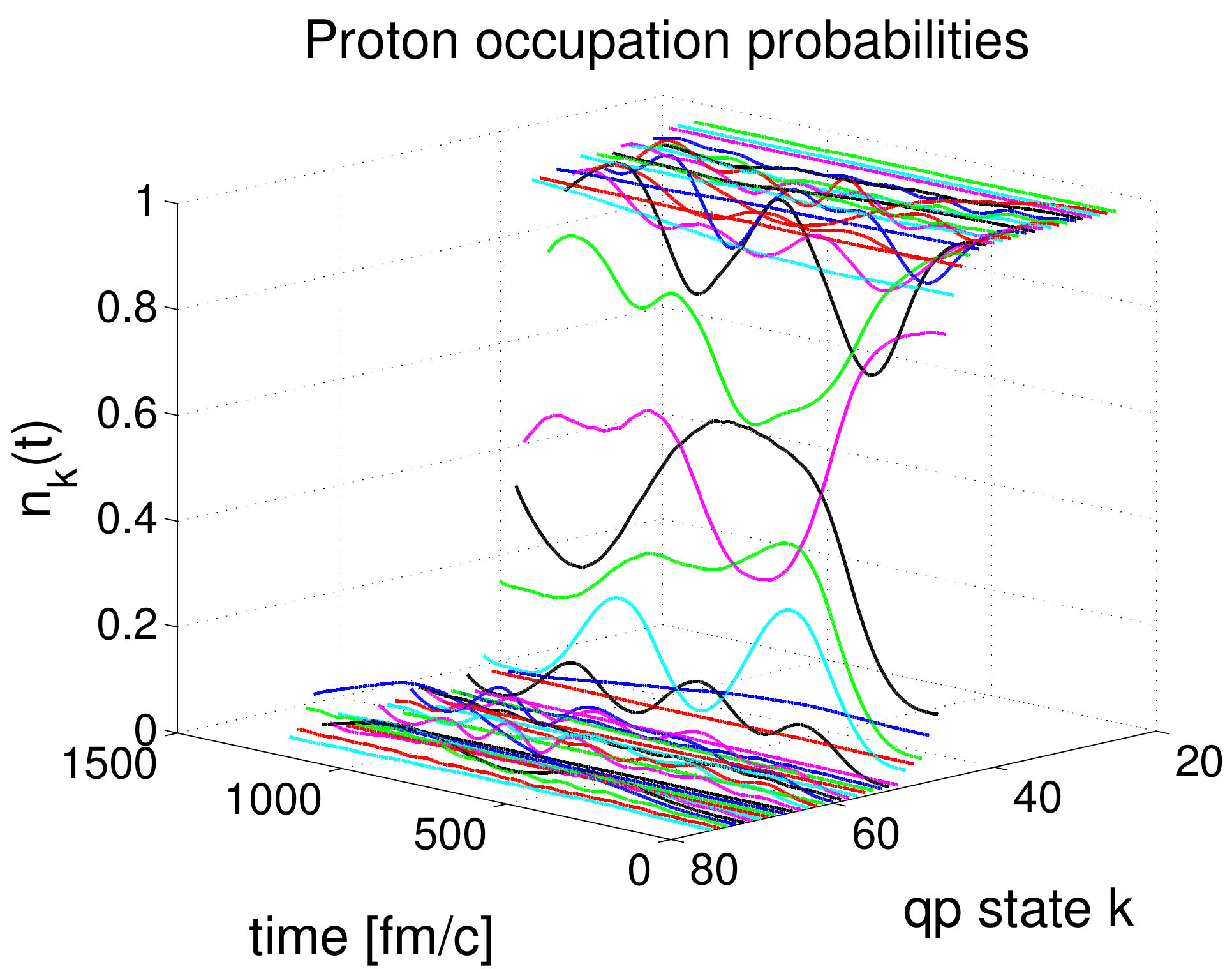}
\includegraphics*[width=0.45\textwidth]{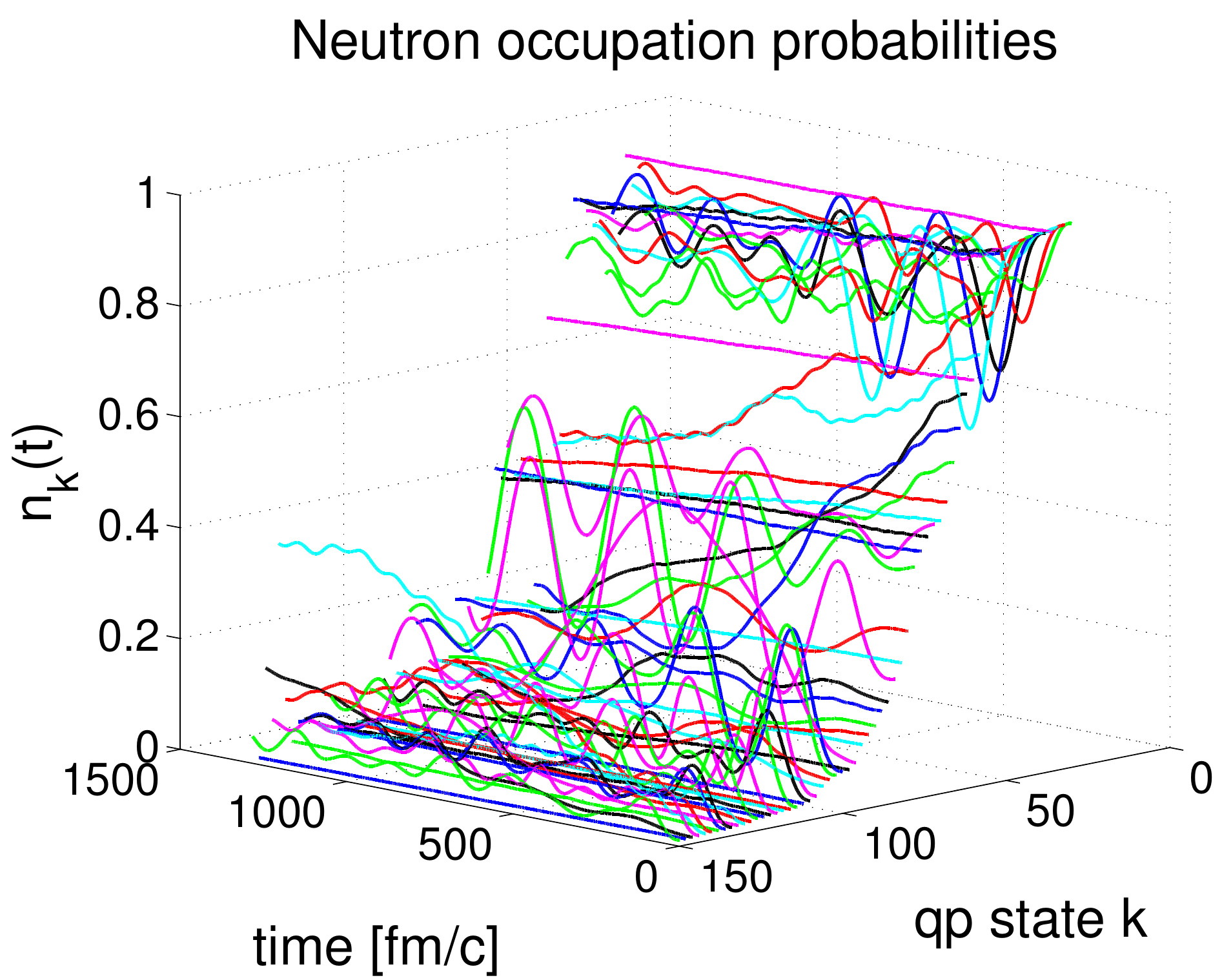}
\caption{ (Color online) The time-dependent proton and neutron occupation probabilities of a set of quasiparticle states around the Fermi level for $^{238}$U calculated as described in the main text with an approximate energy density functional using the SLy4 parametrization. \label{fig1}}
\end{figure}

\begin{figure}
\includegraphics*[width=0.48\textwidth]{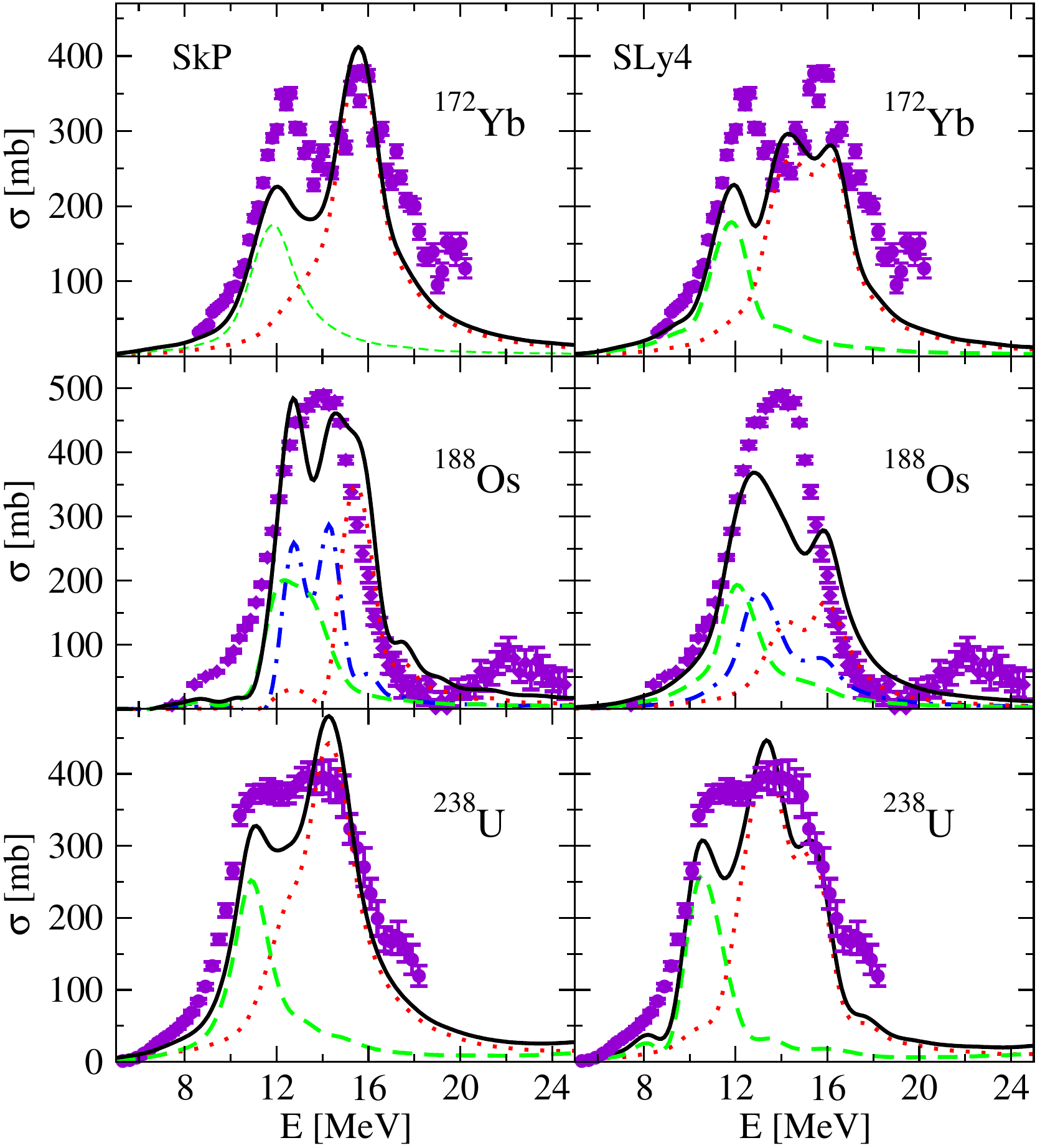}
\caption{ (Color online) The calculated photo-absorption cross-section (solid black line), using two Skyrme force parametrizations for three deformed open-shell nuclei and the experimental $(\gamma ,n)$ cross-sections (solid purple circles with error bars), extracted from Ref. \cite{photo}. With dashed (green), dotted (red) and dotted-dashed (blue) lines we display the contribution to the cross-section arising from exciting the corresponding nucleus along the long axis, the short axis (multiplied by 2 for the prolate nuclei   $^{172}$Yb and $^{238}$U) and the third middle axis in the case of the triaxial  nucleus $^{188}$Os. }
\label{fig:gdr_all}
\end{figure}

From the TDSLDA solutions we construct the occupation probabilities for both proton and neutron quasiparticle states:
\beq
n_k(t)=\sum_{\sigma={\uparrow,\downarrow}}\int d^3r |v_{k,\sigma}({\bf r}t)|^2, \nn
\eeq 
where $k$ labels the proton and neutron quasiparticle wave functions respectively. Because of the complexity of the QRPA equations two types of approximations have been often used in the past in the literature. In one kind of approximation the proton and neutron densities are constructed from 
\beq
v_k({\bf r},\sigma,t)=\sqrt{n_k(t)}\phi_k({\bf r},\sigma,t), \nn
\eeq
where $\phi_k({\bf r},\sigma,t)$ satisfy TDHF-like equations. The occupation probabilities $n_k(t)$ are obtained either in a TDBCS formalism, or are simply assumed to be frozen at their ground state values. One can easily show that in the case of TDBCS the continuity equation is not satisfied \cite{george}, and thus particle number conservation is violated. The assumption that occupation probabilities are frozen to their ground-state values is violated strongly as well, as the two plots in Fig. \ref{fig1} here amply demonstrate. A typical  example of this frozen occupation probabilities approximate approach is in Ref.  \cite{pg}, in which one can find a calculation of the GDR in the triaxial nucleus $^{188}$Os with SLy6, which is very similar to SLy4. The shape of the GDR  line shape calculated in Ref. \cite{pg} is noticeably different from both the experimental one and our results; see Fig. \ref{fig:gdr_all}. We notice that there are experimental indications that a pygmy resonance exists in the nucleus $^{172}$Yb \cite{pigmy}.  We did not find any evidence of a pygmy resonance using either of the two Skyrme energy-density functional parametrizations SLy4 and SkP, and neither did Terasaki and Engel using SkM* \cite{terasaki}.  In  Ref. \cite{terasaki} pairing correlations were treated using a sharp cutoff, a prescription that shows a strong dependence on the cutoff energy, at least in the case of low-lying states \cite{terasaki2}. The approach adopted by us is free of such cutoff dependency in the pairing channel, once the cutoff energy is chosen roughly above 50 MeV \cite{by1}.

In Fig. \ref{fig:gdr_all} we show the results of our calculations and compare them with experimental data \cite{photo}. For the axially deformed nuclei, we performed two calculations perturbing the systems along the longer and the shorter axis, respectively. The triaxial nucleus $^{188}$Os required three runs accordingly. The experimental data include several effects: the coupling of the collective strength to the $1p-1h$  states, especially to the low-lying collective vibrations, Landau damping ($\Gamma_L\approx \hbar v_F/R$, where $v_F$ is the Fermi velocity and $R$ the nuclear radius), particle escape into the continuum ($\Gamma^\uparrow$), and coupling to more complex states ($\Gamma^\downarrow$).  The theoretical Thomas-Reiche-Kuhn sum rule for these nuclei are: 2.49 ($^{172}$Yb), 2.81 ($^{188}$Os), and 3.38 ($^{238}$U) in  MeV$\cdot$barn respectively \cite{ring}. The corresponding energy integrated photo-absorption cross-sections in the interval 8- to 20-MeV are:  2.47, 2.73, and 3.06 in experiment, 1.94, 2.26, and 2.49 for SkP, and 1.84, 2.01, and 2.35 for SLy4 respectively. Both SkP and SLy4 forces underestimate the energy integrated cross-section in this energy interval, even though both Skyrme parametrizations adequately reproduce the average peak position and even the width, in spite of great difference in the isoscalar nucleon effective mass $m^*_{is}=m$  and $m^*_{is}=0.7\,m$, but very similar isovector effective mass, $m^*_{iv}=0.74\,m$ and $m^*_{iv}=0.8\,m$ for SKP and SLy4 respectively. The use of periodic boundary conditions   forced us to use $|{\bf k}|=2\pi/L$ in $F_\tau(x) =  N_\tau\sin( {\bf k}\cdot{\bf r} _\tau )/|{\bf k}| $, instead of $|{\bf k}|\rightarrow 0$, which leads to an underestimate of the cross-sections by $\approx 25\%$. With this trivial correction (obtained by calculating the energy weighted sum rule for the external probe we obtain 2.63(6) ($^{172}$Yb), 3.11(7) ($^{188}$Os), and 3.55(8) ($^{238}$U) for SkP and 2.50(5) ($^{172}$Yb), 2.77(5) ($^{188}$Os), and 3.35(8) ($^{238}$U),  with the SLy4 results being in slightly better agreement with experimental results. The mixing with more complex configurations  ($\Gamma^\downarrow$),  will lead to some depletion of the transition strength out of this energy interval.


We thank G.F. Bertsch and T. Nakatsukasa for discussions and P.-G. Reinhard for literature leads. This work was supported by U.S. DOE Grants No. DE-FG02-97ER41014, No. DE-FC02-07ER41457, and No. DE-AC05-760RL01830 (KJR), and the Polish Ministry of Science under Contract No. N202 128439.  Calculations have been performed on University of Washington Hyak cluster  (NSF MRI Grant No. PHY-0922770), Franklin (Cray XT4, NERSC, DOE Grant No. B-AC02-05CH11231), and JaguarPF (Cray XT5, NCCS, DOE Grant No. DE-AC05-00OR22725). 



\begin{thebibliography}{99}
\bibitem{csb} A term of the type $C[\rho](\rho_p-\rho_n)(|\nu_p|^2-|\nu_n|^2)$, where $\rho_\tau$ and $\nu_\tau$ are the normal and anomalous densities ($\tau=n\,p$) and $C[\rho]$ may depend on $\rho=\rho_p+\rho_n$, will not break charge symmetry.
\end{thebibliography}
\end{document}